\documentclass[twoside,12pt]{article}
\usepackage{epsfig}
\usepackage{amsmath}

\def\Journal#1#2#3#4{{#1} {#2} (#4) #3 }

\def\NPA{{\em Nucl. Phys.} A}

\def\PREP{\em Phys. Rep.}

\def\PRD{{\em Phys. Rev.} D}
\def\PRC{{\em Phys. Rev.} C}

\newcommand{\be}{\begin{equation}}
\newcommand{\ee}{\end{equation}}
\newcommand{\bea}{\begin{eqnarray}}
\newcommand{\eea}{\end{eqnarray}}

\begin{document}

\title{ \vspace{1cm} Relativistic effects in neutrino-Fermi gas interactions}
\author{K.\ Vantournhout, N.\ Jachowicz, and J.\ Ryckebusch\\
\\
Department of Subatomic and Radiation Physics, Ghent University, Belgium}
\maketitle
\begin{abstract} 
We study neutrino interactions in a hadron gas within a relativistic framework.
The hadron matter is described by a non-interacting Fermi gas in beta equilibrium.  We show that the introduction of relativistic effects causes a sizable enhancement of the neutrino-scattering cross sections. \end{abstract}
Neutrino interactions are pivotal in the dynamics of core-collapse supernova explosions and in the cooling of a newly formed neutron star.
In recent years, several studies of neutrino scattering in nucleon matter at supra-nuclear densities have been made \cite{reddy,burrows,horowitz}. Thereby the influence of various aspects of the nuclear dynamics and of the weak nucleon current was investigated.
In this work, we focus on the impact of the implementation of relativistic effects on the description of the process.  

The differential cross section for the interaction of a neutrino $\nu(p_{\nu})$ with a hadron $h_1(p_{h_1})$ in the relativistic gas can be written as
\begin{align}
\mathcal{N} d^{6}\sigma &= \sideset{}{}\sum_{\substack{\alpha_\nu,\alpha_{h_1},\\ \alpha_\ell,\alpha_{h_2}}}\int F^{h_1}_{\alpha_{h_1}}(\vec{p}_{h_1})d^3\vec{p}_{h_1}\:\delta^4(p_\nu+p_{h_1}-p_\ell - p_{h_2})\:\frac{(2\pi)^{10}}{v_{rel}}\:\left|M(\nu,h_1\rightarrow \ell,h_2)\right|^2 \notag \\
& \times (1-F^{h_2}_{\alpha_{h_2}}(\vec{p}_{h_2}))d^3\vec{p}_{h_2} (1-F^{\ell}_{\alpha_\ell}(\vec{p}_\ell)) d^3\vec{p}_\ell, \label{eq1}
\end{align}
where the reaction products are denoted by $\ell(p_{\ell})$ for the outgoing lepton, and $h_2(p_2)$ for the final nucleon, $v_{rel}$ is the relative velocity of the incident particles.   The probability distributions are represented by the Fermi distributions $F^a_{\alpha_a}(\vec{p}_a)$, the quantum numbers ${\alpha_a}$ identify the state  particle $a$ is occupying.
  The Fermi distributions are related through the restrictions  imposed 
by the beta-equilibrium conditions for the $n$, $p$ and $e^-$ in the gas.
 The dynamics of the interaction is contained in the matrix element $M(\nu,h_1\rightarrow \ell,h_2)$ that is calculated in first-order perturbation theory, using the full expression for the hadron vertex-function as given by \cite{llewellyn}.  The normalization factor $\mathcal{N}$ is determined such  that Eq.~(\ref{eq1}) represents the scattering cross-section per nucleon.\\

When considering the effects of relativity, our study shows  that the larger differences are  caused by the relativistic description of the Fermi distribution and not by the implementation of relativity in the dynamics of the neutrino-nucleon interaction.  
For higher momenta, the relativistic Fermi distribution $F(p) = \left[\exp((E(p)-\mu)/kT)+1\right]^{-1}$ obtains a slightly larger weight than the non-relativistic one.  This is illustrated  in the left panel of Fig.~1, showing that especially the tail of the distribution is affected 
by the difference between the relativistic $E(p)$=$\sqrt{p^2+m^2}$ and non-relativistic $E(p)=p^2/2m$ energy expressions.
\begin{figure}[tb]
\begin{center}
\begin{minipage}[t]{16 cm}
\epsfig{file=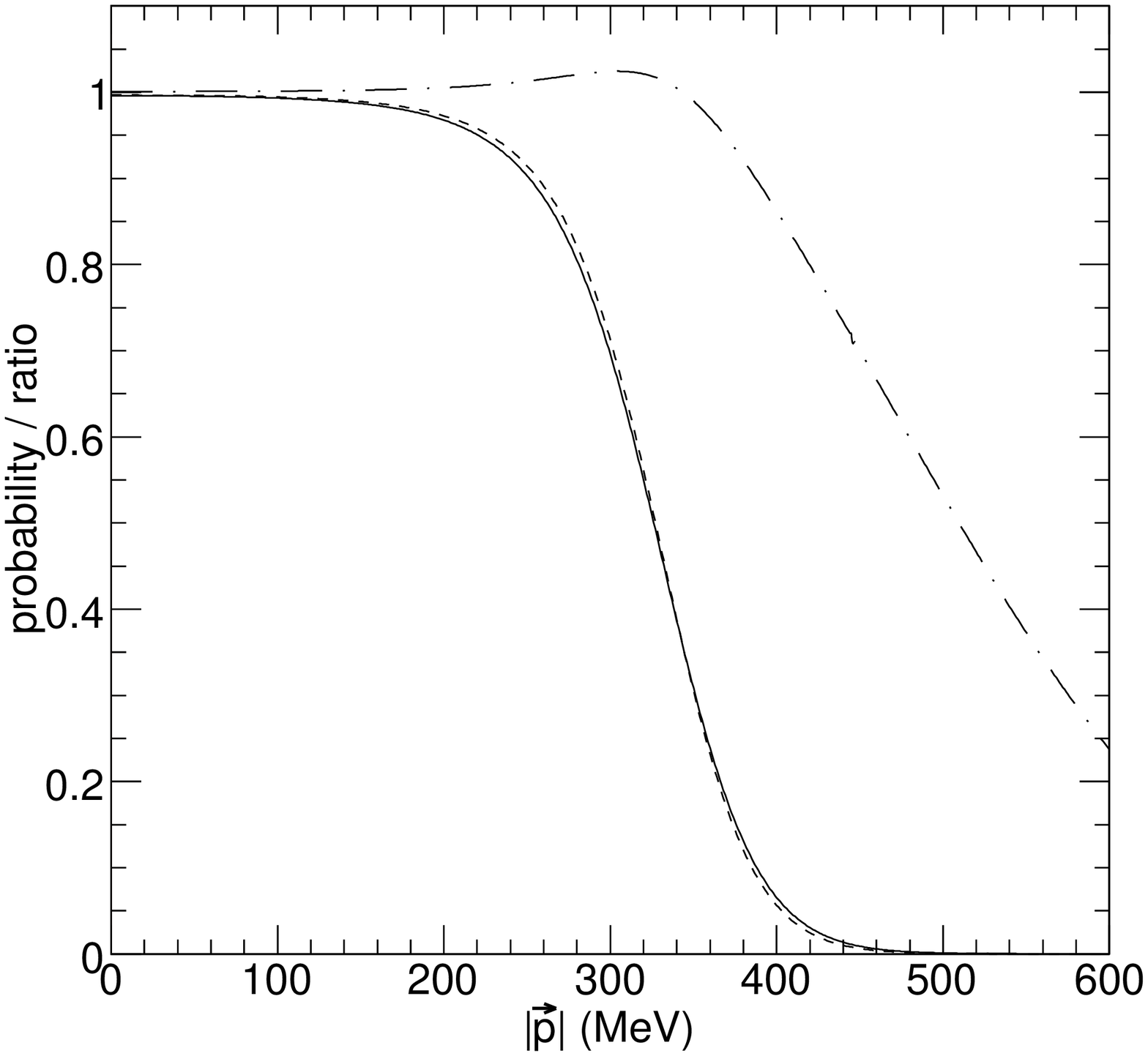,scale=0.4}\epsfig{file=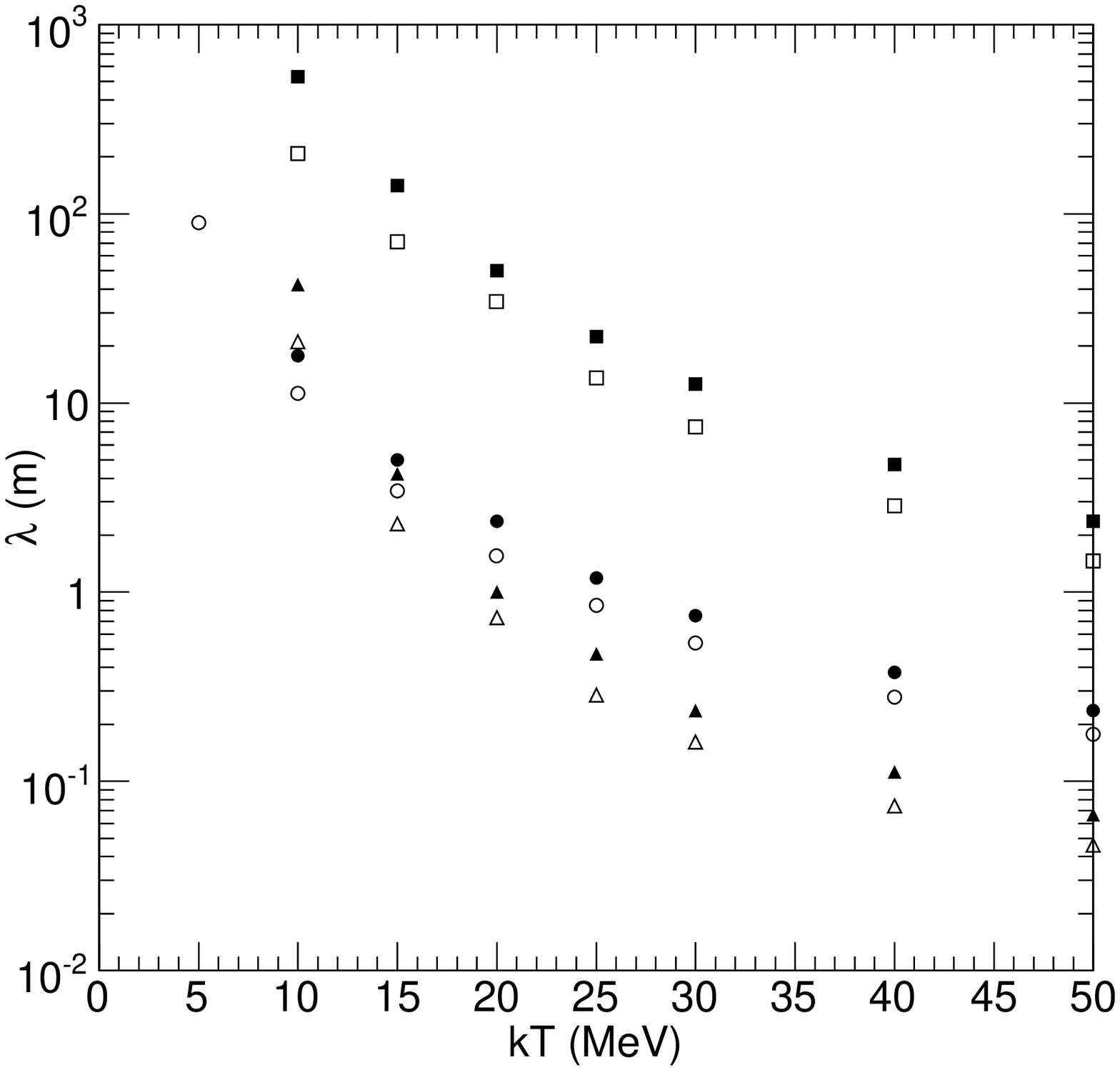,scale=0.4}
\end{minipage}
\begin{minipage}[t]{16.5 cm}
\caption{The left panel compares a relativistic Fermi distribution (solid line) with a  non-relativistic one (dashed) at $kT=10$ MeV, $n=0.16$ fm$^{-3}$. The ratio of both is represented by the  dashed-dotted curve. The right panel illustrates the impact of these differences  on the  neutrino mean free path $\lambda=\frac{1}{n\sigma}$ for nucleon matter in beta equilibrium at density $n$.  The open symbols correspond to our  relativistic calculation,  the full ones compare  them to  the non-relativistic calculation of Ref.~\cite{reddy}.  Triangles represent charged-current neutrino processes, squares (circles) represent neutral-current scattering off  protons (neutrons).  The energy of the incident neutrino was taken to be three times the temperature of the nucleon matter.\label{fig1}}
\end{minipage}
\end{center}
\end{figure}

These relativistic effects are carried over into the calculation of cross sections for interactions in the gas and thus have an important impact on neutrino opacities.
Although the differences in the energy distributions are relatively small, the energy sensitivity of the cross sections (1), rising fast with increasing energies, ensures that the relativistic effects have a sizable influence on the interactions under study.  Relativistic cross sections are generally larger than their non-relativistic counterparts.
  The right panel of Fig.~1 indeed shows  that the opacities obtained within a relativistic calculation are considerably smaller than the ones of the non-relativistic study of Ref.~\cite{reddy}.  Further work on the influence of relativistic effects,  and on the implementation of correlations in the cross section calculation
is in progress \cite{vantournhout}.


\end{document}